\documentclass[preprint2]{aastex}
\newcommand{\SNRrx}{SNR RX J1713.7-3946 (G347.3-0.5)}
\newcommand{\SNRvela}{SNR RX J0852.0-4622}
\newcommand{\Tstart}{t_0}
\newcommand{\Vmax}{V_{\mathrm{max}}^{\mathrm{ej}}}
\newcommand{\fPL}{f_\mathrm{PL}(p)}
\newcommand{\EcrEsn}{E_\mathrm{CR}/E_\mathrm{SN}}
\newcommand{\gamray}{$\gamma$-ray}
\newcommand{\gamrays}{$\gamma$-rays}
\newcommand{\NL}{nonlinear}
\newcommand{\muG}{$\mu$G}
\newcommand{\epRel}{(e/p)_{\mathrm{rel}}}
\newcommand{\Self}{self-similar}
\newcommand{\etamfp}{\eta_\mathrm{mfp}}
\newcommand{\CD}{contact discontinuity}
\newcommand{\tShock}{t_\mathrm{shock}}
\newcommand{\Rsk}{R_{\mathrm{sk}}}
\newcommand{\Vsk}{V_{\mathrm{sk}}}
\newcommand{\pmax}{p_\mathrm{max}}
\newcommand{\pinj}{p_\mathrm{inj}}
\newcommand{\vinj}{v_\mathrm{inj}}

\newcommand{\TP}{test-particle}
\newcommand{\SNa}{SNe}
\newcommand{\SNe}{SNe}

\newcommand{\RFS}{R_\mathrm{FS}}
\newcommand{\RCD}{R_\mathrm{CD}}
\newcommand{\fsk}{f_\mathrm{sk}}
\newcommand{\ffp}{f_\mathrm{p}(p)}
\newcommand{\ffe}{f_\mathrm{e}(p,0)}
\newcommand{\ffeZ}{f_\mathrm{e}(p,z)}
\newcommand{\LD}{L_D(p)}
\newcommand{\rg}{r_g}

\newcommand\Rtot{r_\mathrm{tot}}
\newcommand\Rsub{r_\mathrm{sub}}

\newcommand\Twind{T_\mathrm{w}}
\newcommand\rhoEj{\rho_\mathrm{ej}}
\newcommand\rhoWind{\rho_\mathrm{w}}
\newcommand{\xx}[1]{\times 10^{#1}}

\newcommand{\myemail}{\texttt{don\_ellison@ncsu.edu}}
\newcommand{\gccemail}{\texttt{chenai@physics.rutgers.edu}}
\newcommand{\rel}{relativistic}

\newcommand{\nonrel}{non-relativistic}
\newcommand{\RfsRcd}{R_\mathrm{FS}/R_\mathrm{CD}}

\newcommand{\syn}{synchrotron}
\newcommand{\synch}{synchrotron}
\newcommand{\brem}{bremsstrahlung}

\newcommand{\IC}{inverse-Compton}
\newcommand{\pion}{pion-decay}
\newcommand\tSNR{t_\mathrm{SNR}}
\newcommand\EnSN{E_\mathrm{SN}}
\newcommand\Mej{M_\mathrm{ej}}
\newcommand\etainj{\eta_\mathrm{inj}}
\newcommand{\kmps}{km s$^{-1}$}

\newcommand\Msun{\mathrm{M}_{\odot}}
\newcommand{\vWind}{v_\mathrm{w}}
\newcommand{\sigwind}{\sigma_\mathrm{w}}
\newcommand{\Mdot}{dM/dt}

\newcount\listnorom
\newcommand\listromanDE{\global\advance \listnorom by 1
{\lowercase\expandafter{(\romannumeral\listnorom)}\ }}
\newcommand\newlistroman{\listnorom=0}

\slugcomment{Submitted to ApJ April 2005, Accepted June 2005}

\shorttitle{Radio and X-ray Profiles in SNRs}
\shortauthors{Ellison and Cassam-Chena{\"\i}}

\begin{document}

\title{Radio and X-ray Profiles in Supernova Remnants Undergoing Efficient Cosmic Ray
    Production}

\author{Donald C. Ellison }
\affil{Physics Department, North Carolina State University, Box 8202, Raleigh, NC 27695}
\email{\myemail}

\author{Gamil Cassam-Chena{\"\i}}
\affil{Department of Physics and Astronomy, Rutgers University,
136 Frelinghuysen Rd, Piscataway NJ 08854-8019}
\email{\gccemail}

\author{Submitted to ApJ April 2005, accepted June 2005}

\begin{abstract}
The strong shocks in young supernova remnants (SNRs) should accelerate cosmic rays
(CRs) and no doubt exists that \rel\ electrons are produced in SNRs. However, direct
and convincing evidence that SNRs produce CR nuclei has not yet been obtained and
may, in fact, be long in coming if current \gamray\ observatories do not see an
unambiguous \pion\ feature. Nevertheless, the lack of an observed \pion\ feature does
not necessarily mean that CR ions are not abundantly produced since ions do not
radiate efficiently.
If CR ions are produced efficiently by diffusive shock acceleration (DSA), their
presence will modify the hydrodynamics of the SNR and produce morphological effects
which can be clearly seen in radiation produced by electrons.
We describe in some detail our CR-hydro model, which couples DSA with the remnant
hydrodynamics, and the \syn\ emission expected for two distinct parameter sets
representing type Ia and type II supernovae.
Several morphological features emerge in radial profiles, including the forward shock
precursor, which are observable with current X-ray observatories and which should
definitively show if young SNRs produce CR ions efficiently or not. For the specific
case of SN 1006, we conclude, as have others, that the extremely short X-ray scale
heights observed near the outer shock argue convincingly for the efficient production
of CR ions.
\end{abstract}


\section{INTRODUCTION}
Collisionless shocks in supernova remnants (SNRs) are believed to produce the
majority of Galactic cosmic rays (CRs), at least up to the so-called ``knee'' near
$10^{15}$~eV \citep[see][for a recent review]{Hillas2005}.  While there is little
doubt from the \syn\ interpretation of radio observations that young SNRs produce GeV
electrons, and this is probably true for TeV electrons as well from the
interpretation of nonthermal X-rays, there is as yet no unambiguous direct evidence
that SNRs produce \rel\ ions.
This is somewhat paradoxical considering that the observed electron to proton ratio
in CRs is $\sim 0.01$ and virtually all models of diffusive particle acceleration in
collisionless shocks, the most cited mechanism for producing CRs, predict that ions
receive far more energy than electrons \citep[see, for example,][and references
therein]{BaringEtal99}. Relativistic electrons, of course, radiate far more
efficiently than do ions, leaving open the possibility that a large majority of the
energy in \rel\ particles in SNRs lies in hard to see ions.
In this paper, we model SNR evolution coupled with the efficient production of CRs
\citep[our so-called CR-hydro model, e.g.,][]{EDB2004} and make a number of
predictions for the \syn\ emission from electrons which will be influenced by the
presence of otherwise unseen \rel\ ions.
For a recent summary of observations and models of \syn\ emission
in SNRs, see \citet{CassamEtal2005} which addresses many of the
issues discussed here using a self-similar approach. 

In order to power CRs, the shocks in SNRs must be capable of
placing $\sim 10$\% of the supernova (SN) explosion energy into
\rel\ ions over their lifetime \citep[e.g.,][]{Drury83,BE87}. In
fact, the strong shocks in young SNRs may be far more efficient
than this \citep[e.g.,][]{Ellison2000,HRD2000,DEB2000} and place
enough energy in \rel\ particles so that nonlinear feedback
effects modify the shock structure, the evolution of the remnant,
and the radiative properties \citep[e.g.,][]{BEK96,DEB2000}. 
As we show below, structural changes produced by DSA translate into changes in \syn\
emission that are large enough to be investigated with modern, high spatial
resolution, radio and X-ray observatories.
In particular, we calculate the \syn\ emission profiles for typical shell-type Ia and
II supernova parameters and show how these profiles provide important constraints on
the underlying particle acceleration mechanism and magnetic field structure.

Particle acceleration 
influences the SNR {evolution}
because \rel\ particles
produce less pressure for a given energy density than do \nonrel\
particles.\footnote{This follows since the ratio of specific heats, $\Gamma$,
decreases as particles become \rel\ and the pressure $P=(\Gamma -1)e$, where $e$ is
the energy density.}
Therefore, when \rel\ particles are produced and/or energetic particles escape from
the shock system, the shocked gas becomes more compressible,
i.e., it acts as if it has a softer equation of state and the remnant hydrodynamics
are modified.
The softer effective equation of state means that compression ratios well in excess
of four can be produced in non-radiative, collisionless shocks
\citep[e.g.,][]{BE87,JE91}, and since the energy going into \rel\ particles is drawn
from the shock-heated thermal population, the temperature of the shocked gas can be
much less than that expected from \TP\ shock acceleration
\citep[e.g.,][]{Ellison2000,DEB2000}.
In addition to modifying the 
{evolution}
and the temperature of the shocked gas,
changes in the compression of the fluid should result in changes in the compression
of the magnetic field implying that \syn\ emission from \rel\ electrons will
vary strongly with the efficiency of DSA and the orientation and strength of the
magnetic field.

Perhaps the most important morphological aspect of this CR-hydro coupling is that the
ratio of the forward shock radius, $\RFS$, to the radius of the \CD, $\RCD$, may be
much less than in the \TP\ case \citep[see][]{DEB2000,EDB2004}.
If, as is generally believed, shocks put far more energy into accelerated ions than
electrons, it is the efficient production of cosmic ray {\it ions} that reduces
$\RfsRcd$ from \TP\ values. However, since the interaction region between the forward
shock (FS) and the \CD\ (CD) can be sometimes estimated or determined with modern
X-ray telescopes (SN 1006 is an example where the CD is not seen), radiating
electrons can reveal the presence of these otherwise unseen \rel\ ions.

Another clear morphological prediction from efficient DSA discussed below is that
radial profiles of X-ray emission will be strongly peaked and form sheet-like
structures at the FS. This effect comes largely from the large shock compression
ratios which compress the magnetic field behind the FS and result in severe radiative
losses for electrons producing X-rays. Without efficient particle acceleration, the
radial profiles of X-rays will be smoother and more closely resemble those for radio
emission.

In addition to the radio and X-ray profiles in the interaction region between the CD
and the FS, we calculate the emission in the FS precursor. We show that the structure
of the X-ray precursor depends strongly on assumptions made for the magnetic field
compressibility. If the magnetic field is compressed substantially at the FS, as is
likely, the ratio of X-ray intensity immediately upstream of the shock to that at the
FS drops dramatically. In this case, line-of-sight projection effects produce
profiles that are fully consistent with the extremely short scale heights seen in SN
1006 by \citet{BambaEtal2003} or \citet{LongEtal2003}, even though TeV electrons with
long diffusion lengths are present.
We conclude for this particular remnant, as did \citet{BKV2003}, that CR ions are
being efficiently produced and their presence is revealed by radiating electrons.
{We note that the strong magnetic fields we describe at the FS are produced by
compression not from magnetic field amplification resulting from cosmic-ray streaming
instabilities, such as predicted by \citet{BL2001}. Magnetic field amplification at
the FS is not included in our model.}

\section{CR-HYDRO MODEL}

We calculate the hydrodynamic evolution of a SNR coupled to efficient DSA with a
radially symmetric model described in detail in \citet{EDB2004} and references
therein. We do not consider CR production at the reverse shock since we assume the
magnetic field in the ejecta is the frozen-in field from the SN progenitor and, as
such, will be too small to produce significant particle acceleration or non-thermal
emission without large enhancement factors \citep[see][for a discussion of efficient
DSA at reverse SNR shocks]{EDB2005}.

Any realistic model of a SNR will have several parameters for both the environment
and the physical processes controlling the evolution and particle acceleration. Here,
we concentrate on changes in the SNR evolution and emission produced by CR
production, and choose two fairly distinct models as prototypes, one with parameters
typical of type Ia \SNa\ and the other with parameters likely those of type II
\SNa. 
These models differ by the initial density profile in the ejecta\footnote{Since we
don't consider acceleration at the reverse shock, the different ejecta composition in
type Ia and type II \SNe\ is not important. For a discussion of how composition might
influence DSA, see \citet{EDB2005}.}
and the density and magnetic field profiles in the ambient medium.
Within these models, we investigate the effects of varying the CR production
efficiency and the magnetic field structure.

\subsection{Type Ia prototype}
For our type Ia prototype, we assume the density profile of the ejecta material is
exponential \citep{DC98}, the total ejecta mass is $\Mej=1.4\,\Msun$, the explosion
energy is $\EnSN=10^{51}$ erg, and a uniform ambient medium density, $n_p$, with a
temperature of $T_0 = 10^4$ K. Here, $n_p$, is the proton number density and we
assume there is an additional 10\% contribution of helium nuclei.
We assume the magnetic field in the interstellar medium (ISM), $B_0$, is also
constant and take $B_0=10^{-5}$ G as a default value.  We typically view our type Ia
models at an age $\tSNR=400$ yr, similar to the age of Tycho's SNR, when the shock
speed is roughly $4000$\,\kmps.

\subsection{Type II prototype}
For our type II prototype, we assume the initial density profile of the ejecta
material is a power law in radius, $\rhoEj \propto r^{-n}$, with a constant density
plateau region at small radii \citep[e.g.,][]{Arnett88}. We take $n=9$ in all of our
type II models. For the total ejecta mass we take $\Mej=2\,\Msun$,
and the explosion
energy is set to $\EnSN=3\xx{51}$ erg \citep{LH03, CO03}. The density of the pre-SN
wind is taken as $\rhoWind = A r^{-2}$, where $A=\Mdot/(4 \, \pi \, \vWind)$, $\Mdot$
is the mass loss rate, and $\vWind$ is the wind speed (both assumed constant).  We
use typical values $\vWind=20$ \kmps, $\Mdot = 2\xx{-5}\,\Msun$ yr$^{-1}$
\citep{CO03}, and take a constant wind temperature $\Twind=10^4$ K.

Following \cite{CL94}, we assume the unshocked magnetic field in the
pre-SN wind is
\begin{equation} \label{Bwind}
B_{0}(r) = \left( \sigwind \, \vWind \, \Mdot \right)^{1/2} /
\, r \ ,
\end{equation}
or
\begin{eqnarray}
B_{0}(r) &=& 2.6 \left( \frac{\sigwind}{0.1}  \right)^{1/2}
\left(  \frac{\vWind}{10 \, \mathrm{km/s}}  \right)^{1/2}\,
\\ \nonumber & \times & \left( \frac{\Mdot}{10^{-5} \, \mathrm{M}_{\odot}\, \mathrm{/
yr}}
 \right)^{1/2} \, \left( \frac{r}{1 \, \mathrm{pc}} \right)^{-1}\,
\mu\mathrm{G}
\ ,
\end{eqnarray}
where $\sigwind$ is the constant ratio of magnetic field energy density to
kinetic energy density in the wind.
This expression assumes that the magnetic field is frozen in the constant stellar
wind and is only valid in the equatorial plane for distances $r$, much greater than
the radius of the pre-SN star.\footnote{ {Equation~(\ref{Bwind}) only applies if
the forward shock has not reached the stellar wind termination shock. We assume the
forward shock is within the bubble in all of the examples discussed here. } }
Off the plane, $B(r)$ will fall off more rapidly than $1/r$,
but we ignore this effect in our spherically symmetric models. The
value of $\sigwind$ for stars other than the sun is not well known
but, for concreteness, we take $\sigwind=0.1$.
We typically view our type II models at $\tSNR=400$ yr, to match our type Ia models
and for comparison to SNR Cassiopeia A (Cas A), when the shock speed is roughly
$6000$\,\kmps.

\subsection{Acceleration model}
For the diffusive shock acceleration process, we use the algebraic model of
\citet{BE99} and \citet{EBB2000} where the injection efficiency is parameterized and
the superthermal spectrum, $f(p)$, is a broken power law, $\fPL$, with an exponential
turnover at high momenta, $f(p) \propto \fPL \exp{(-p/\pmax)}$.
The algebraic model solves the \NL\ DSA problem at each time step of the hydro
simulation given the shock speed, shock radius, ambient density and temperature, and
ambient magnetic field determined in the simulation. With the accelerated
distribution, an effective ratio of specific heats is calculated and used in the
hydrodynamic equations, completing the coupling between the two \citep[see][for a
full discussion]{EDB2004}.

The injection parameter,
$\etainj$, is the fraction of total protons injected into the DSA process and values
$\etainj \gtrsim 10^{-4}$ typically yield efficient particle acceleration rates where
$10 \%$ to $99\%$ of the available energy flux goes into \rel\ protons.\footnote{Note
the difference between the fraction of protons injected into the acceleration
process, $\etainj$, and the acceleration efficiency.  The acceleration efficiency is
the fraction of total energy flux going into \rel\ particles including all ions and
electrons.  Given $\etainj$ and the other shock parameters, the electron spectrum is
determined with two additional parameters, the electron to proton ratio at \rel\
energies, $\epRel$, and the electron to proton temperature ratio immediately behind
the shock, $T_e/T_p$ \citep[see][for a full discussion]{EBB2000}.}
The maximum momentum, $\pmax$, is determined by setting the acceleration time equal
to the SNR age $\tSNR$ or, by setting the diffusion length of the highest energy
particles equal to some fraction, $\fsk$, of the shock radius $\Rsk$, whichever gives
the lowest $\pmax$ \citep[see, for example,][]{BaringEtal99}.
In all of the models presented here we take $\fsk=0.05$.  We assume Bohm diffusion so
that the scattering mean free path, $\lambda$, is on the order of the gyroradius,
$r_g$, i.e., $\lambda$ = $\etamfp r_g$ with $\etamfp= 1$ and $\rg=pc/(qB)$.  Here,
$p$ and $q$ are the particle momentum and charge, respectively, $B$ is the magnetic
field at the acceleration site, and $c$ is the speed of light.
Note that while our estimate of $\pmax$ requires a specific assumption for the mean
free path, the acceleration model itself only assumes that the scattering mean free
path is a strongly increasing function of momentum.  In the absence of radiative
losses, the maximum kinetic energy particles receive in DSA depends only on the
particle charge and $\pmax$ is the same for protons and electrons as long as both
are \rel.

\subsection{Synchrotron emission and losses}

As the forward shock overtakes fresh ambient medium material, the shock accelerates
these particles and produces a nonthermal distribution as described in detail in
\citet{EDB2004} and \citet{EDB2005}.\footnote{We ignore pre-existing CRs and inject
and accelerate only thermal particles over taken by the shock.}
Once the particle distribution is produced in a shell of material at the shock, it is
assumed to remain in that shell as the shell convects and evolves behind the shock.
During the evolution, particles experience adiabatic and \syn\ losses and these
losses are calculated as in \citet{Reynolds98}.

In calculating the \syn\ emission and losses, we evolve the magnetic field as
described, for example, in \citet{RC81} or \citet{Reynolds98}.  Consider a fluid
element which is now at position $r$ with density $\rho(r)$.
At an earlier time, this fluid element was shocked at a position $r_i$ where the
density immediately behind the shock was $\rho_2$. The radial and tangential
components of the field immediately behind the shock at $r_i$, were $B_{2r}$ and
$B_{2t}$, respectively.
If the magnetic flux is frozen in the fluid, the field at the downstream position,
$r$, is given by
\begin{equation} \label{Bevo}
B^2(r) = B^2_{2r} \left ( \frac{r_i}{r} \right )^4 +
B^2_{2t} \left ( \frac{\rho(r)}{\rho_2} \right )^2
\left ( \frac{r}{r_i} \right )^2
\ .
\end{equation}
For the magnetic field configuration across the shock, we assume either that
$B_2=B_0$, as in a parallel shock, or that the field is fully turbulent upstream and,
following \citet{VBKR2002}, set the immediate downstream magnetic field
\begin{equation} \label{B_comp}
B_2= \sqrt{1/3 + 2 \Rtot^2/3}~B_0
\ ,
\end{equation}
where $\Rtot$ is the shock compression ratio.\footnote{Here and
elsewhere the subscript 0 (2) indicates values immediately ahead of (behind) the
shock.}
{Note that $B_2$ does not include any amplification effects such as described by
  \citet{BL2001}.} 
Using $B(r)$ obtained in eq.~(\ref{Bevo}), the evolution of the electron distribution
under combined adiabatic and \syn\ losses is calculated and, at the end of the
simulation, the \syn\ emission in each shell is determined as in
\citet{BaringEtal99}.\footnote{In calculating electrons losses, we include \IC\
losses off the cosmic microwave background radiation as described in
\citet{BaringEtal99}. For protons, radiative losses are unimportant for typical SNR
magnetic fields.}

\begin{figure}        
\epsscale{.99} \plotone{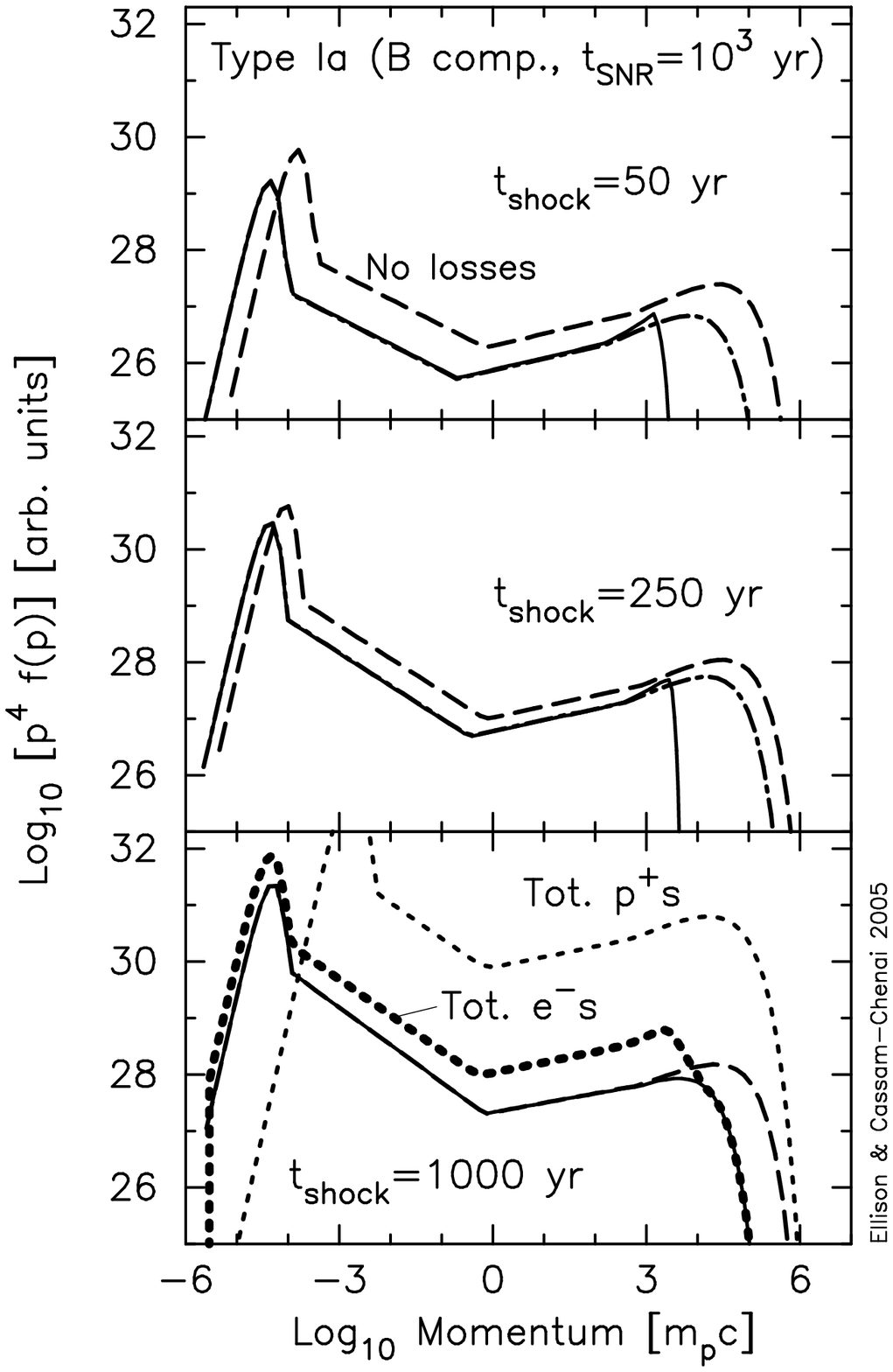}
\caption{Phase space electron distributions multiplied by $p^4$. In all panels, the
  dashed curves were produced, without losses, in shells at the FS at the times
  indicated, $\tShock$.  The solid curves are these distributions at $\tSNR=10^{3}$
  yr with adiabatic and \syn\ losses taken into account and, in the top two panels,
  the dot-dashed curves are these distributions at $\tSNR=10^{3}$ yr with only
  adiabatic losses included. The heavy (light) dotted curve in the bottom panel is
  the total electron (proton) distribution, with all losses, from the interaction
  region between the CD and the FS at $\tSNR$. The magnetic field is compressed at
  the shock as in eq.~(\ref{B_comp}) and $\etainj=10^{-3}$.
\label{fp_No_loss}}
\end{figure}

In Fig.~\ref{fp_No_loss} we show electron momentum phase-space distribution
functions, $f(p)$, for a type Ia SNR model discussed more fully in
Section~\ref{results} below.  In each panel, the dashed curve is the distribution
calculated immediately after production at the age indicated (i.e., at $\tShock$) and
the solid curve is this distribution at the end of the simulation (i.e., at
$\tSNR=1000$ yr) after experiencing adiabatic and \syn\ losses. In the top two
panels, the dot-dashed curves show the electron distribution at $\tSNR=1000$ yr when
only adiabatic losses are included.

The shock accelerated distribution, before losses, is a broken power law above a
thermal distribution with an exponential cutoff at the maximum momentum \citep[i.e.,
eq.~12 in][with $\alpha=1$]{EBB2000}.
Adiabatic losses affect all particles
(shifting the entire distribution to lower momenta, i.e., $p
\propto \rho^{1/3}$), while \syn\ 
influence mainly the
highest energy electrons. For the parameters of this model, the
highest momentum electrons accelerated at early times are strongly
depleted and a distinct \syn\ bump is observed just below the
sharp maximum momentum cutoff.
The heavy-weight dotted curve in the bottom panel of Fig.~\ref{fp_No_loss} is the
electron distribution at the end of the simulation summed over the interaction region
between the \CD\ and the forward shock. For comparison we show with the light-weight
dotted curve the summed proton distribution at the end of the simulation. For this
example, the electron to proton ratio at \rel\ energies, $\epRel$, is set to 0.01,
similar to that of Galactic cosmic rays, and the electron to proton temperature ratio
immediately behind the shock, $T_e/T_p$, is set to 1 \citep[see][for fuller
discussion of these parameters]{EBB2000}.  The difference between the electron and
proton spectra in the bottom panel of Fig.~\ref{fp_No_loss} illustrates how DSA
typically puts far more energy into protons than electrons.

\subsection{Upstream precursor}
{The algebraic acceleration model of \citet{BE99} doesn't explicitly include the
  geometry of the shock precursor. However, we can estimate the precursor upstream of
  the forward shock in the following way.}
{At any particular time,}
the proton distribution in the outer most shell, $\ffp$, produces the precursor.
We assume that the protons of momentum $p$ in this distribution ``feel'' a flow speed
$u(z)$ and magnetic field $B(z)$, where $z$ is the diffusion length, $\LD$, measured
upstream from the FS.  The diffusion length $\LD = \kappa(p)/u(z)$, where
$\kappa=\lambda v/3$ is the diffusion coefficient, $v$ is the particle speed, and
$u(z)$ is the flow speed at $z$ measured in a frame at rest with the shock.

We use information from $\ffp$ to estimate $u(z)$ and $B(z)$ and obtain
$\LD$. Because of shock smoothing, the compression ratio in the FS that produced
$\ffp$ ranges from the subshock compression, $\Rsub$, felt by protons with the
superthermal injection momentum $\pinj$, to the overall compression, $\Rtot$, felt by
protons with $\pmax$. Intermediate values of compression, $r(p)$, felt by protons or
electrons with momentum $p$ between $\pinj$ and $\pmax$, can be estimated with a
linear extrapolation between $r(p)$ and $\log{(pv)}$, i.e.,
\begin{equation} \label{eff_comp}
r(p) = \Rsub + G(p) \; (\Rtot - \Rsub) \ ,
\end{equation}
where $p v$ is proportional to the diffusion length and
\begin{equation}
G(p) = \frac{\log{(pv)} - \log{(pv)_\mathrm{inj}}}
         {\log{(pv)_\mathrm{max}} - \log{(pv)_\mathrm{inj}}}
\ .
\end{equation}
Here $(pv)_\mathrm{max}= \pmax \, c$, $(pv)_\mathrm{inj}= \pinj \, \vinj$, and
$\vinj$ is the particle speed corresponding to $\pinj$. Note that since $\pinj$ and
$\etainj$ combine to give a single free injection parameter, the specific value of
$\pinj$ is unimportant for the results discussed here \citep[see][for recent work on
injection in a semi-analytic, nonlinear DSA model]{BGV2005}.

With equation~(\ref{eff_comp}), we estimate the flow speed felt by a particle with
momentum $p$ as
\begin{equation}
u(z) = \Vsk \frac{r(p)}{\Rtot}
\ ,
\end{equation}
and the magnetic field felt by this particle is either
\begin{equation} \label{noBcomp}
B(z) = B_0
\end{equation}
or
\begin{equation} \label{Bcomp}
B(z) = B_0 \sqrt{ \frac{1}{3} + \frac{2}{3} \left ( \frac{\Rtot}{r(p)} \right )^2}
\ ,
\end{equation}
depending on if the magnetic field is compressed in the precursor (as in
Eq.~\ref{B_comp}) or not. Here $\Vsk$ is the forward shock speed in the rest frame of
the SN.

Given $u(z)$ and $B(z)$, the diffusion length of an electron can be determined and,
in a fashion similar to \citet{Reynolds98}, we assume that
electrons of momentum $p$ are distributed upstream from the shock
such that
\begin{equation}
\ffeZ = \ffe \exp{ \{ -z[1/\LD + 1/(\fsk \RFS)] \} }\ ,
\label{fp_prec}
\end{equation}
where $\ffe$ is the electron distribution in the outer most shell ($z=0$) at the end
of the simulation and $\fsk \RFS$ sets the distance ahead of the shock where
particles freely leave the system. The electron distribution, $\ffe$, contains the
effects of \syn\ and \IC\ losses which occur during acceleration.

\begin{figure}         
\epsscale{.99} \plotone{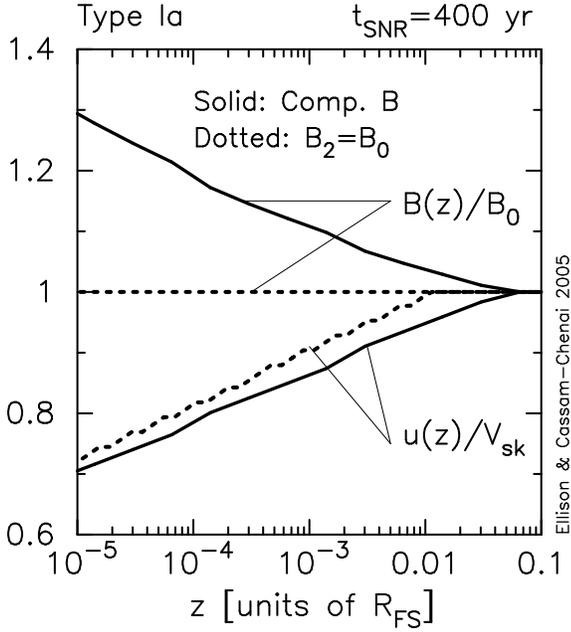} 
\caption{Precursor flow speed $u$ and magnetic
field, $B$, as functions of $z$, the distance upstream from the forward shock. The
magnetic field is in units of $B_0$, the far upstream value, and $u(z)$ is in units
of the forward shock speed, $\Vsk$, measured in the rest frame of the SN. For the
type Ia models shown here, we take $B_0=10$\,\muG. For our type II wind models (not
shown), we assume that the length scale of the wind is large compared to the
precursor scale so that $B(z)$ in eq.~(\ref{Bcomp}) is obtained using a constant
$B_0$, where $B_0$ is the immediate upstream field at $\tSNR$.
\label{precursor}}
\end{figure}

The above relations are approximations in that they ignore the precise form for the
smooth precursor flow speed. However, we have verified that the precursor emission is
relatively insensitive to this smoothing and that our approximations
adequately describe the spatial dependencies important for
predicting the \syn\ precursor.  Typical results are shown in
Fig.~\ref{precursor} where the solid curves are for compressed $B$
and the dotted curves are for uncompressed $B$.

\section{RESULTS} \label{results}
\subsection{Radial emission}
Using the parameters for our type Ia prototype, we plot in Fig.~\ref{Ia_stack} the
\syn\ emission as a function of radius for one radio (1-1.4 GHz; solid curves) and
two X-ray bands (0.1-1 keV dashed curves; 1-10 keV dotted curves).  We present four
models, two with $\etainj=10^{-3}$, which produces very efficient DSA with nearly
100\% of the energy flux crossing the shock going into \rel\ particles, and two with
$\etainj=10^{-5}$, which yields essentially a \TP\ result with less than 1\% of the
energy flux going into CRs and where the influence of shock accelerated protons on
the hydrodynamics is small. For each $\etainj$ we show a case with a compressed field
(labelled B comp.) and one with uncompressed field either in the shock or the
precursor (labelled $B_2=B_0$). In the compressed field case, we assume, as in
\citet{BKV2002}, that the magnetic field is fully turbulent upstream of the shock and
is compressed in the precursor as described by equation~(\ref{Bcomp}).
The curves are normalized to each energy band's flux at the forward
shock.\footnote{The results of the CR-hydro model, at early times, depend on the
initial conditions which, unavoidably, are somewhat arbitrary. The initial
conditions, in turn, influence the emission at the CD seen in Figs.~\ref{Ia_stack}
and \ref{II_stack_10yr}.  For all of the results presented here, the simulation is
started at a time $\Tstart=10$\,yr with the initial ejecta speed varying linearly
with radius from zero to a maximum speed $\Vmax=0.1c$.  The initial maximum radius of
the ejecta is set by $\Vmax$ and $\Tstart$ and the early stages of the simulation,
and therefore the \syn\ emission at the CD, depend on $\Vmax$ and $\Tstart$.  Of
course, the later evolution of the SNR is nearly independent of the starting
conditions, as long as the total kinetic energy and ejecta mass stay the same. Since
the X-ray emission is dominated by losses at the CD, it is only the radio emission at
the CD that depends strongly on $\Vmax$ and $\Tstart$. For a full discussion of the
start up conditions for the CR-hydro model, see \citet{EDB2004}.}

\newlistroman

Fig.~\ref{II_stack_10yr} shows similar results for our type II prototype where, as in
Fig.~\ref{Ia_stack}, the emission is viewed at $\tSNR=400$ yr.

Comparing these figures, we note the following:

\listromanDE The two SN types have very similar profiles 
at least for the parameters
used here.
One noticeable difference occurs for the $\etainj =10^{-5}$ cases where the 
type II radio profiles are flatter than the type Ia profiles.
Later, in association with Fig.~\ref{Ia_II_inj}, we show in more detail
that changes in $\etainj$ and other parameters influence the
SN types rather differently  and may offer help in distinguishing
the types.

\begin{figure}         
\epsscale{.99}
\plotone{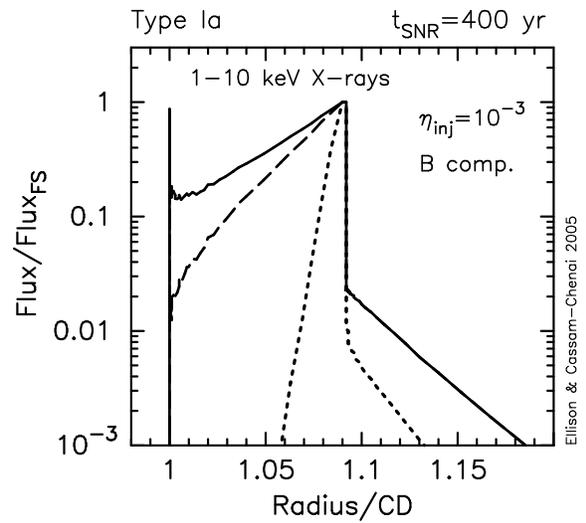}
\caption{Radial \syn\ emission for 1--10 keV X-rays. The dotted curve shows the
  profile with both adiabatic and \syn\ losses included, the dashed curve has only
  adiabatic losses, and the solid curve was calculated with no losses. Since there
  are no adiabatic losses in the precursor, the dashed and solid curves are identical
  in the precursor. 
\label{Xray_No_syn}}
\end{figure}

\listromanDE In the interaction region between the contact discontinuity and the
forward shock, the X-ray \syn\ falls off more rapidly than the radio emission.
As mentioned in discussing Fig.~\ref{fp_No_loss}, the electrons producing the
radio emission suffer only adiabatic losses, while the higher energy electrons
producing the X-rays suffer adiabatic losses combined with \syn\ and \IC\ losses.
In Fig.~\ref{Xray_No_syn} we show profiles for the 1--10 keV band with no losses
(solid curve), with just adiabatic losses (dashed curve), and with adiabatic plus
radiative losses (dotted curve).
Since, for typical SNR parameters, the nonthermal X-ray emission comes from the
exponential part of the electron spectrum, the X-ray emission will be extremely
sensitive to changes in the spectrum coming from any type of loss mechanism.

\listromanDE The radio emission can have a secondary peak at the
CD, while the X-ray emission, with \syn\ losses, always drops
precipitously at the CD. As just mentioned, the radio emission at
the CD is sensitive to the starting conditions of the hydro model
but, in any case, the secondary peak is less noticeable 
in projection as we show below.

\listromanDE With efficient DSA and a compressed magnetic field (top panels of
Figs.~\ref{Ia_stack} and \ref{II_stack_10yr}), the X-ray fall-off is extremely rapid
and the X-ray emission can appear as an extremely thin sheet at the FS.

\listromanDE The precursor outside of the FS falls slowly if the magnetic field is
not compressed at the shock, but drops sharply immediately upstream of the shock when
$B$ is compressed, with or without efficient DSA (top two panels in
Figs.~\ref{Ia_stack} and \ref{II_stack_10yr}).  The sharp drop due to the compressed
field will make the X-ray precursor faint and difficult to detect compared to the
emission at the FS. Without compression, the precursor should be observable,
providing an important diagnostic for the magnetic
field structure. Note that the radio precursor has an extremely short upstream
diffusion length for all cases and will not be detectable if the diffusive length
scale is anywhere near as small as we predict.

\listromanDE Comparing the $\etainj=10^{-3}$ panels against the $\etainj=10^{-5}$
panels in Figs.~\ref{Ia_stack} and \ref{II_stack_10yr} shows that the distance
between the CD and the FS is nearly a factor of two greater in the test-particle case
than with efficient DSA.  Since the limit of the shocked ejecta gives an idea of the
position of the CD, $\RfsRcd$ is measurable in several young SNRs with {\it Chandra} and
{\it XMM-Newton}, making this morphological difference a powerful diagnostic for
efficient DSA.

\begin{figure}         
\epsscale{.99}
\plotone{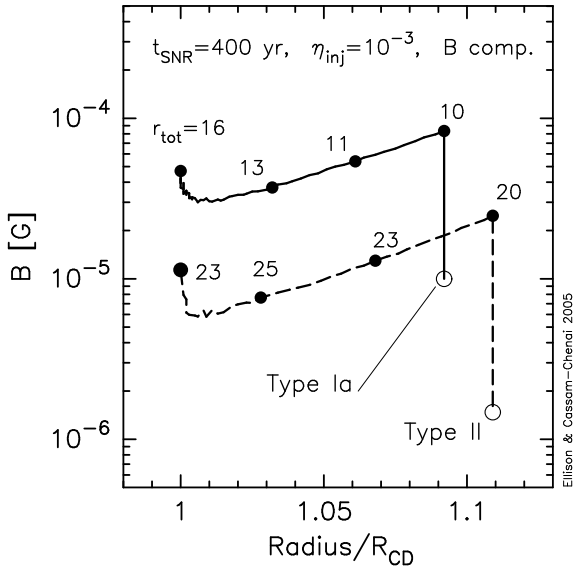}
\caption{Magnetic field strength in the interaction region between the CD and FS for
  a type Ia SNR (solid curve) and a type II SNR (dashed curve). The number above each
  solid dot is the compression ratio that parcel of gas experienced when it was
  shocked. The open circles at the ends of the curves indicate the unshocked field at
  the end of the simulation, i.e., at $\tSNR=400$\,yr.
\label{B_Rtot}}
\end{figure}

In Fig.~\ref{B_Rtot} we show the magnetic field structure, at $\tSNR=400$\,yr, in the
transition region between the CD and FS for our two prototypes with compressed $B$
and $\etainj=10^{-3}$. The numbers at specific points on the curves indicate the
compression ratio, $\Rtot$, at the FS at the time that particular parcel of gas was
shocked. It is notable that $\Rtot \gg 4$ in all cases. The difference in $\Rtot$
between the two models comes about  mainly from the lower magnetic field in the pre-SN
wind for the type II model which results in larger compression ratios.
The end of the curves, marked with an open circle, show the immediate upstream,
unshocked magnetic field, $B_0$, at $\tSNR$.  For type Ia, $B_0=10$\,\muG\ and is
independent of time, while for type II, $B_0(r)$ falls off with radius as in
equation~(\ref{Bwind}) and at $\tSNR=400$\,yr is $\simeq 1.5$\,\muG.  A thorough
discussion of the influence magnetic field strength has on $\Rtot$ is given in
\citet{EDB2005}.

\subsection{Line-of-sight projections}
In Fig.~\ref{Ia_LOS} we show line-of-sight projections for some of the results shown
in Fig.~\ref{Ia_stack}.  Even in projection, it is clear that the radio emission
falls off less rapidly behind the FS than the X-ray emission.  Projection has little
effect on the upstream precursor so the large differences seen in Fig.~\ref{Ia_stack}
with and without magnetic field compression are similar in projection.
The decrease in $\RfsRcd$ for efficient particle acceleration is less obvious in
projection but, since the CD generally shows up via thermal X-ray emission, $\RfsRcd$
remains an important diagnostic for the presence of efficient CR ion acceleration.
Line-of-sight projections of the results shown in Fig.~\ref{II_stack_10yr} are
similar.

An important feature that is in the line-of-sight projections and not in the radial
profiles is the offset of radio and X-ray peaks at the FS. In Fig.~\ref{Ia_LOS_FS},
the projections for the type Ia models of Fig.~\ref{Ia_stack} with compressed
magnetic fields are plotted as a fraction of the FS radius. With or without efficient
DSA, the radio peak (solid curve) occurs inside the X-ray peaks.  Behavior such as
this is observed in several SNRs including G347 \citep[][]{Lazendic2004}, Kepler
\citep[][]{DeLaneyEtal2002}, Tycho \citep[][]{DecourchelleEtal2001}, and Cas A
\citep[][]{LongEtal2003}.  We note, however, that there is another radio peak
coincident with the X-ray peak in Tycho \citep[e.g.,][]{DickelEtal91}.
For the efficient acceleration case (top panel), the two X-ray
peaks are also well separated.
Note also that because of projection effects, the maximum emission occurs inside of
the FS.
As emphasized by \citet{BKV2003}, care must be taken not to interpret the peak
emission as the position of the FS, as done by \citet{BambaEtal2003} for SN 1006.
The actual upstream precursor is indicated in Fig.~\ref{Ia_LOS_FS} with a ``P.''

In Fig.~\ref{Ia_II_inj} we compare the line-of-sight 1-10 keV X-ray projections for
both type Ia and type II prototypes calculated with different DSA injection
efficiencies.
While the absolute normalization is arbitrary, the curves show the correct relative
normalization between the various models and, as expected, the \TP\ cases with
$\etainj=10^{-5}$ have lower absolute emissivities.
In both panels, the solid curves have $\etainj=10^{-3}$, the dashed curves have
$\etainj=10^{-4}$, the dotted curves have $\etainj=10^{-5}$, and all models have
magnetic field compression (note the different vertical scales in the two panels).
For both SN types, the ratio $\RfsRcd$ increases noticeably as the acceleration
becomes less efficient, but $\RfsRcd$ increases somewhat more rapidly for type II
SNRs.
Also, for both SN types, the morphology of the X-ray emission varies strongly with
$\etainj$: for efficient DSA, there is a pronounced peak at the rim, while the emission
is much broader for inefficient DSA. This difference offers another important
diagnostic for efficient DSA.

In Fig.~\ref{Mdot} we keep all parameters of our $\etainj=10^{-3}$ type II model
constant except the wind speed, $\vWind$, and the mass loss rate, $\Mdot$.  In the
top panel, $\vWind=20$\,\kmps\ and $\Mdot$ varies, as indicated,
and the light-weight dashed curve has $\etainj=10^{-4}$; all other curves in
Fig.~\ref{Mdot} have $\etainj=10^{-3}$.  
As $\Mdot$ increases, there is an increase in $\RfsRcd$ indicating, among other
things, that self-similarity is no longer a good approximation at $\tSNR=400$\,yr.
In the bottom panel, $\Mdot = 2\xx{-5}\,\Msun$ yr$^{-1}$ and $\vWind$ is varied as
indicated.  Now, the profiles are relatively insensitive to the changes in $\vWind$,
suggesting that self-similarity does apply.

In considering Figs.~\ref{Ia_II_inj} and \ref{Mdot} it's important to note that while
$\RfsRcd$ is reduced substantially with efficient CR production in type Ia SNRs,
values of $\RfsRcd > 1.3$ can occur in type II SNRs with very efficient DSA.
The acceleration efficiency for the $\etainj=10^{-4}$ model in Fig.~\ref{Mdot}
(light-weight dashed curve) is
greater than 50\% over most of its 400\,yr lifetime.  This may be relevant for
remnants like Cas A and 1E0102.2-7219 which show $\RfsRcd \sim 1.4$.

\subsection{Radio emission vs. ejecta profile and age}
It is well known that young SNRs with power-law ejecta and power-law ambient medium
density profiles have self-similar solutions if CR production is absent or
unimportant \citep[i.e.,][]{ch82,Chev82Let}. This will be true for the efficient
production of CRs as well if the CR production is time invariant \citep[][]{ch83}.
If nonlinear DSA occurs and the acceleration efficiency varies with time, the
self-similarity is broken \citep[see][]{EDB2004}, as is the case with an exponential
ejecta density distribution \citep[e.g.,][]{Dwarkadas2000}, or for a power-law ejecta
distribution once the reverse shock enters the plateau region of the ejecta.

\begin{figure}         
\epsscale{.99} \plotone{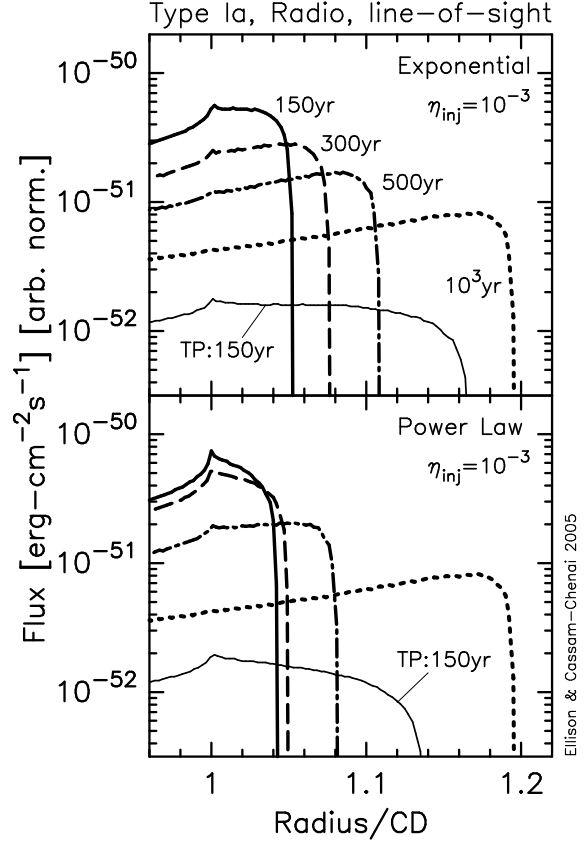} 
\caption{Radio \syn\ profiles for our type
Ia prototype at various ages, as indicated, and for an exponential ejecta
distribution (top panel) and a power law ejecta distribution with $n=9$ (bottom
panel).  The line styles indicate the same ages in both panels and in all cases,
except for the curves marked TP:150\,yr,
$\etainj=10^{-3}$ and the ambient magnetic field is $B_0=10$\,\muG.  As shown by the
solid curve in the bottom panel, a power law ejecta distribution produces a radio
profile at early times that peaks near the CD.  The light-weight solid curves are
\TP\ results shown for comparison.
\label{radio_age}}
\end{figure}

In Fig.~\ref{radio_age} we show radio emission profiles at various $\tSNR$ for type Ia
models with $\etainj=10^{-3}$ having exponential (top panel) and power law (bottom
panel) ejecta density profiles. In \Self\ evolution, the ratio $\RfsRcd$ remains
constant and this is approximately the case for a power-law ejecta density profile
for $\tSNR \lesssim 300$ yr. At later times, the self-similarity is broken, as is
the case at all times for exponential ejecta density profiles. The light-weight solid
curves are \TP\ profiles at 150 yr for comparison.

Besides $\RfsRcd$, the structure of the radio emission in the interaction region
between the CD and the FS depends on the assumed ejecta distribution and on the age
of the SNR. At early times for the power-law case (solid curve, bottom panel of
Fig.~\ref{radio_age}), the radio emission peaks near the \CD. This result is
consistent with the self-similar model described in \citet{CassamEtal2005} but, as
discussed above, depends somewhat on the starting conditions of the CR-hydro model.
At later times the emission drops inside the FS and, as expected, the details of the
ejecta profile cease to be important.
The curves for 1-10 keV X-rays are not shown, but due to radiative losses and
contrary to the radio, they peak strongly just behind the FS for all $\tSNR$ as shown
in Fig.~\ref{Ia_LOS_FS}.

\subsection{Acceleration efficiency}
\begin{figure}         
\epsscale{.99} \plotone{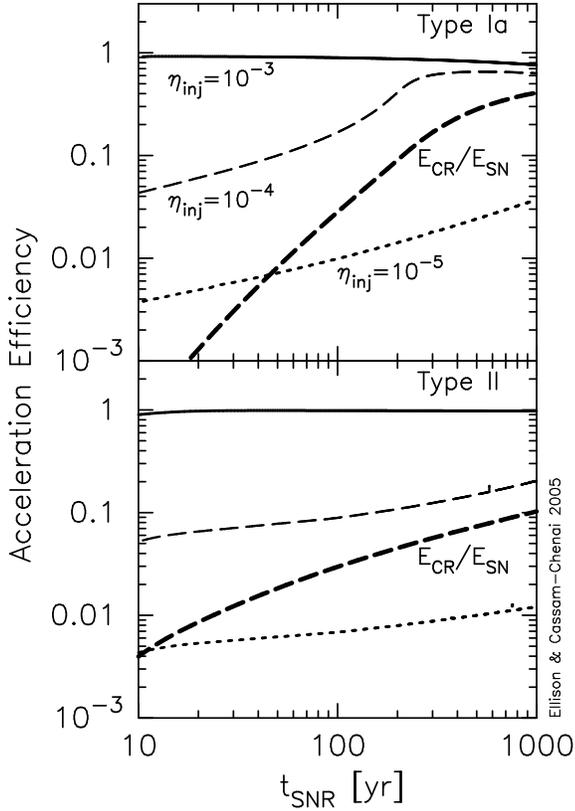} 
\caption{The light-weight curves are the
shock acceleration efficiencies for our two SN prototypes for various $\etainj$, as
indicated. The heavy-weight dashed curves in both panels are the fractions of total
SN explosion energy going into CRs for the case where $\etainj=10^{-4}$. The line
styles indicate the same values of $\etainj$ in both panels.
\label{CR_eff}}
\end{figure}

In Fig.~\ref{CR_eff} we show the acceleration efficiency, i.e., the fraction of
energy flux crossing the shock that goes into \rel\ ions \citep[see eq.~13
of][]{EBB2000}, for various $\etainj$ (light-weight curves) and the fraction of total
SN explosion energy put into CRs, $\EcrEsn$, for $\etainj=10^{-4}$ (heavy-weight
dashed curves). These models use our type Ia and II prototype parameters.
For the, perhaps, extreme case of $\etainj=10^{-3}$, the fraction of bulk flow energy
flux (in the shock rest frame) that is placed in \rel\ ions is $>80\%$ during the
1000 yr span shown for both SNR prototypes.  Even for $\etainj=10^{-4}$, the
efficiency is $> 10\%$ most of the time and more than 10\% of the total SN explosion
energy can be put into CRs over the 1000 yr lifetime.

Of course the actual injection efficiency of SNR shocks is uncertain and, as noted by
\citet{VBK2003}, injection may vary over the surface of the SNR and be significantly
less where the magnetic field is highly oblique \citep[see][for a discussion of
parallel versus oblique shock geometry in SN
1006]{RothenflugEtal2004}. \citet{VBK2003} estimate that to supply the galactic CR
population the overall efficiency need only be $\sim 20$\% of the maximum values
obtained by DSA. \citet{Dorfi90} and \citet{BEK96} obtained similar values.
Nevertheless,
if the shocks in supernova remnants accelerate cosmic ray {\it ions} this efficiently via
diffusive shock acceleration, clear signatures of the
acceleration will be present in the radiation produced by {\it electrons}.

\section{DISCUSSION}
\subsection{Narrow interaction region}
Perhaps the most unambiguous indication of efficient CR production in SNRs is an
interaction region between the \CD\ and the forward shock which is considerably
narrower than predicted without efficient acceleration \citep[e.g.,][]{BE2001}. While
the ratio $\RfsRcd$ depends on various parameters, efficient DSA can easily result in
the FS being less than half the distance ahead of the CD predicted with \TP\
acceleration (see Figs.~\ref{Ia_stack}, \ref{II_stack_10yr}, and \ref{Ia_II_inj}).
This may explain observations of $\RfsRcd$ which are considerably less than the
smallest value predicted by \TP, self-similar models, as is the case for Tycho's
\citep[e.g.,][]{ReynosoEtal1997,DecourchelleEtal2001} and Kepler's
\citep[e.g.,][]{DeLaneyEtal2002,cad04a} SNRs.

Even in SNRs such as Cas A and 1E0102.2-7219 in the Small Magellanic Cloud
\citep[e.g.,][]{GotthelfEtal2001,GaetzEtal2000,HRD2000}, where the FS and CD are well
separated, DSA may be quite efficient. As shown in Figs.~\ref{Ia_II_inj} and
\ref{Mdot}, moderately efficient acceleration and/or the presence of a pre-SN wind
can result in $\RfsRcd \gtrsim 1.3$. Thus, while the observation of $\RfsRcd =
1.0-1.1$ can be explained naturally if CR ions are being produced efficiently in type
Ia \SNa, larger values of $\RfsRcd$ do not necessarily exclude efficient acceleration
but may be representative of type II \SNa\ with pre-SN winds.

\subsection{Precursor and small-scale structure}
In some SNRs extremely small spatial scales in X-ray emission are
observed at the FS.  Using {\it Chandra} observations,
\citet{LongEtal2003} and \citet{BambaEtal2003} have independently
examined emission profiles in several thin filaments in projection
in the northeast shell of SN 1006 which show scale lengths as
short as $0.04$ pc (assuming a distance to the SNR of $\sim 2$\,kpc).

In Fig.~\ref{Bamba} we compare our type Ia prototype model with $\etainj = 10^{-3}$
to the SN 1006 observations.
We represent the observations with dashed lines which roughly span
the maximum and minimum scale heights determined by
\citet{BambaEtal2003} (see their Table~4). 
Even though we have not
attempted a detailed fit to SN 1006, it's clear that our
compressed $B$ model (solid curve) matches the overall
observations quite well and the shortest scale heights are
extremely well modeled.
As emphasized by \citet{BKV2003}, the shortest scale heights occur inside the forward
shock and are produced by projection effects when $B$ is compressed and there is a
sharp drop in emissivity behind the shock. The actual upstream precursor (indicated
with a ``P'' in Fig.~\ref{Bamba}) has a much longer scale height as expected from TeV
electrons but is not easily discernable with {\it Chandra} against background
emission.

While our efficient acceleration model with compressed $B$ fits quite well, our
uncompressed model (dotted curve) clearly does not fit, nor does a \TP\ model (not
shown), as is clear from examining the bottom panel of Fig.~\ref{Ia_LOS}.  As far as
we can tell, our results are in complete agreement with those of \citet{BKV2003}
\citep[see also][]{Ballet2005} and provide convincing evidence for highly compressed
magnetic fields and efficient DSA.

\subsection{Adiabatic and \syn\ losses and the offset of radio and X-ray peaks}
Nonthermal X-ray emission in a fixed energy band is very sensitive to both adiabatic
and radiative losses.  For typical SNR parameters, \syn\ X-rays are produced in large
part by the exponential tail of the electron distribution. Therefore, any energy loss
results in a large drop in emissivity.  This contrasts with the adiabatic losses of
the electrons producing radio emission.
Since radio is produced by lower energy electrons in the power law portion of the
distribution rather than the exponential part, emission in a fixed energy band is
less sensitive to adiabatic losses. If nonlinear effects from efficient DSA are
important, the fixed band radio is even less effected by adiabatic losses since the
portion of the electron distribution producing radio is likely to be concave, i.e.,
flattering with increasing energy.

The \syn\ loss rate will be greater if the magnetic field is compressed at the shock
and, therefore, will depend on the acceleration efficiency. As we show in
\citet{CassamEtal2005} and in  Fig.~\ref{Ia_II_inj} here, the morphology of
the X-ray emission near the FS varies noticeably with $\etainj$, peaking more
strongly as the acceleration efficiency increases since electrons lose energy before
convecting far downstream.  This feature provides an important diagnostic for
acceleration efficiency.

A direct consequence of X-ray emitting electrons suffering more losses than radio
emitting ones, is an offset in the peak emission of the projected flux at the FS. As
shown in Fig.~\ref{Ia_LOS_FS}, the radio emission peaks well within the X-ray
emission.
The separation will depend on the acceleration efficiency since, for a given set of
supernova parameters, models with efficient DSA have larger compression ratios and
larger downstream magnetic fields. The larger the field, the sharper is the drop in
X-ray emission behind the shock, and the closer to the FS position with be the peak
X-ray emission.

\section{CONCLUSIONS}
We have presented a detailed discussion of the influence of efficient diffusive shock
acceleration on the radial profiles of \syn\ emission in young SNRs.
The evidence that collisionless shocks, in general, can accelerate particles with
high efficiency is convincing.  There are direct spacecraft observations confirming
it \citep[e.g.,][]{Eich81,BE87,EMP90,BOEF97,Terasawa99}, plasma simulations show
efficient acceleration consistent with spacecraft observations
\citep[e.g.,][]{STK92,EGBS93,GBSEB97}, Galactic cosmic-ray energetics and composition
suggest it \citep[e.g.,][]{AxfordICRC82,EDM97}, and theoretical models certainly
allow it \citep[e.g.,][]{ALS77,Drury83,EE84,JE91,BEK96,MD2001,KJG2002,Blasi2002}.
An unresolved question, of course, is whether or not shock acceleration is efficient
in SNRs.
If DSA is as efficient in accelerating {\it ions} as suggested, the acceleration
process will be nonlinear and will noticeably modify the SNR structure and
evolution. We have shown for typical type Ia and type II SN parameters that these
structural changes, most important of which is the increased shock compression,
produce clear signatures in the \syn\ radiation emitted by {\it electrons}.
We note, incidentally, that signatures in the thermal emission may also be present
since the energy which goes into \rel\ ions comes out of the bulk thermal plasma and
produces a drastic reduction in the temperature of the shocked gas
\citep[e.g.,][]{DEB2000,HRD2000,EDB2004}.

Of course, our assertion that the \NL\ effects seen in the structure of SNRs are
evidence for the efficient acceleration of ions rather than electrons depends on how
the energy of shock accelerated particles is distributed between electrons and ions.
While no definitive theory exists describing this partition, the source of the energy
going into superthermal particles is the bulk kinetic energy of the converging
upstream and downstream plasmas. 
Diffusive shock acceleration occurs, at its most basic level,
when particles diffuse across the shock and scatter nearly
elastically off the converging plasmas on either side of the shock.  When particles
are accelerated from the thermal background, this process favors heavy
particles and it is generally assumed that shocks put far more energy into ions than
electrons.
There is direct evidence for this disparity in acceleration efficiency at the low
Mach number shocks which have been studied in the heliosphere
\citep[e.g.,][]{Feldman85,Terasawa99} \citep[see also][]{EllisonEtalWorkshop94}, but
there is no direct evidence, one way or the other, in the much stronger shocks which
exist outside of the heliosphere. Nevertheless, with some confidence, we believe the
structural changes we have discussed are produced by ion acceleration with the
radiating electrons being passive markers of the effect.\footnote{We note that
so-called shock surfing has been suggested by a number of workers as an effective way
of transferring shock energy into electrons \citep[see][for example, and references
therein]{HoshinoSurf2002}. A thorough discussion of this mechanism is beyond the
scope of this paper, but we note that while some descriptions of this effect show
large energy gains by electrons, nonlinear effects are almost certain to limit the
effectiveness of this process \citep[see][]{SSM2003}, particularly in the strong
shocks we envision for young SNRs.}

While direct evidence for the production of CR ions in SNRs would be the observation
of a \pion\ spectral feature in GeV--TeV \gamrays, such \gamrays\ are difficult to
detect with the significance necessary to distinguish a \pion\ feature from \IC\ or
\brem\ radiation.  Furthermore, in low density regions, \IC\ may outshine \pion\
emission, leaving the question of CR ion production for these SNRs open regardless of
the sensitivity of \gamray\ telescopes.
{The best chance of seeing a strong \pion\ signal is when a SNR interacts with a
dense medium such as the \synch-dominated \SNRrx\ interacting with molecular clouds
\citep[see][and references therein]{cad04b}.  HESS (High Energy Stereoscopic System)
has recently measured, with high significance, the 1--10 TeV energy spectrum in this
remnant \citep[][]{AharonianNature2004} and in \SNRvela\
\citep[][]{AharonianVela2005} and while \pion\ is certainly the most likely emission
mechanism, it is not possible, based on TeV emission alone, to reliably determine
the different \gamray\ components in these spectra.
It should now be possible to test for \pion\ emission using the morphology since HESS
has, for the first time, produced \gamray\ images of these remnants, and the
morphology of \IC\ and \pion\ should be quite different. }
Observations in the MeV range by GLAST should help significantly to distinguish
\pion\ from lepton emission and may provide incontrovertible evidence for or against
SNRs as the source of CRs ions.

\begin{figure}         
\epsscale{.99}
\plotone{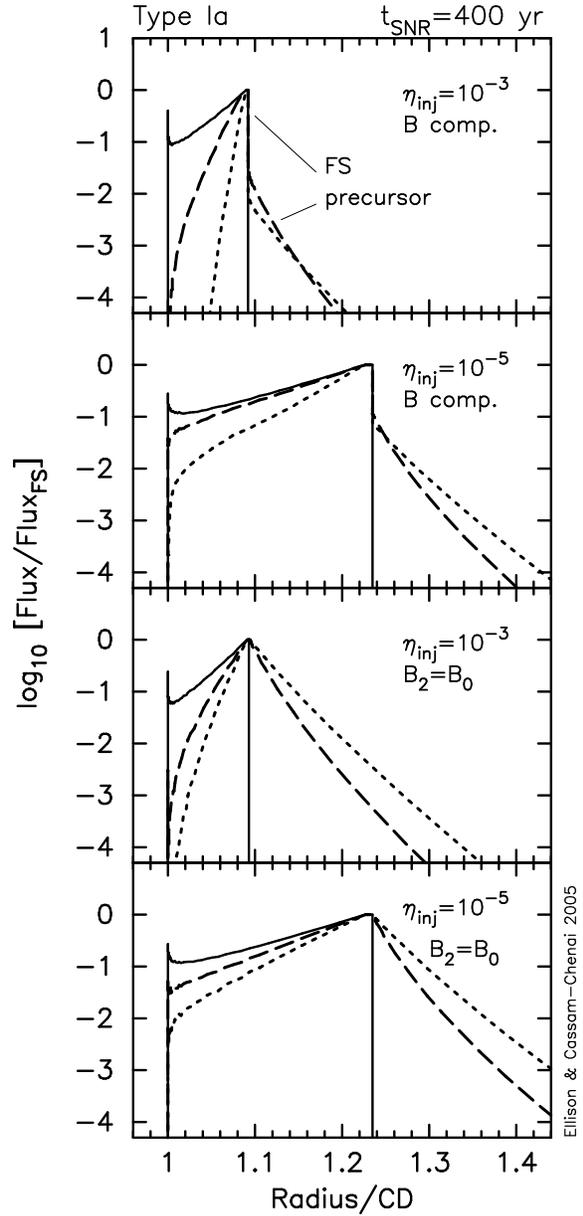}
\caption{Radial \syn\ emission in three energy bands for two magnetic field
  configurations and two shock injection efficiencies, $\etainj$. In all panels, the
  solid curve is radio (1--1.4 GHz), the dashed curve is low energy X-rays
  (0.1--1 keV), and the dotted curve is high energy X-rays (1--10 keV). In the top
  two panels $B$ is compressed as in equation~(\ref{Bcomp}), while in the bottom two
  panels, $B_2=B_0$. The flux of each band is normalized to its value at the FS.
\label{Ia_stack}}
\end{figure}

\begin{figure}         
\epsscale{.99}
\plotone{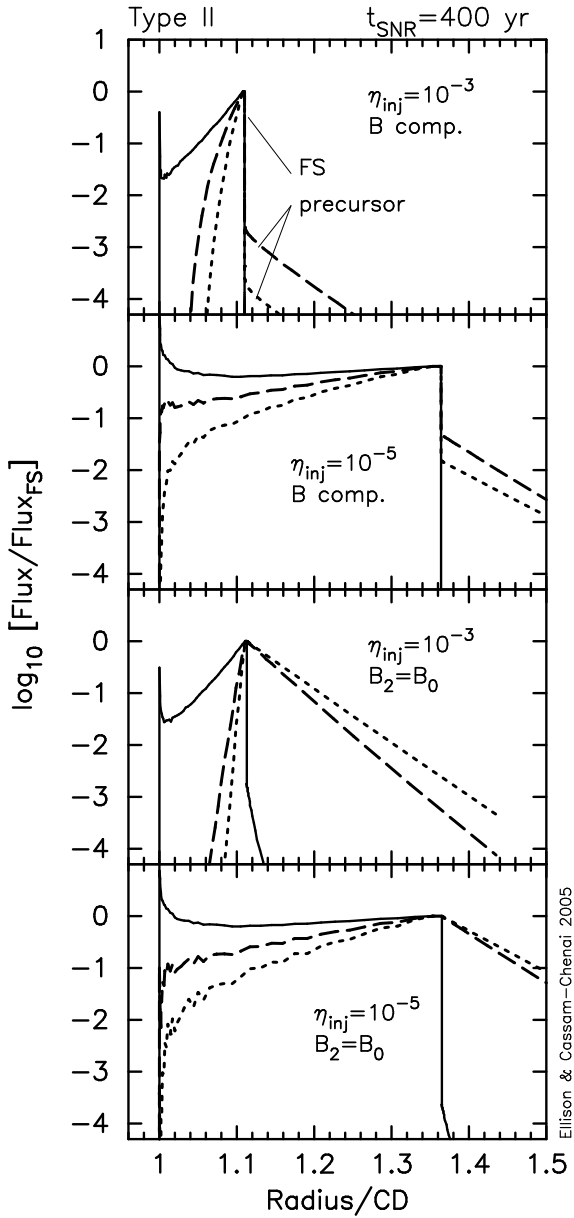}
\caption{Radial \syn\ emission for three energy bands as in Fig.~\ref{Ia_stack} for
  our type II SNR prototype.
\label{II_stack_10yr}}
\end{figure}

\begin{figure}         
\epsscale{0.99}
\plotone{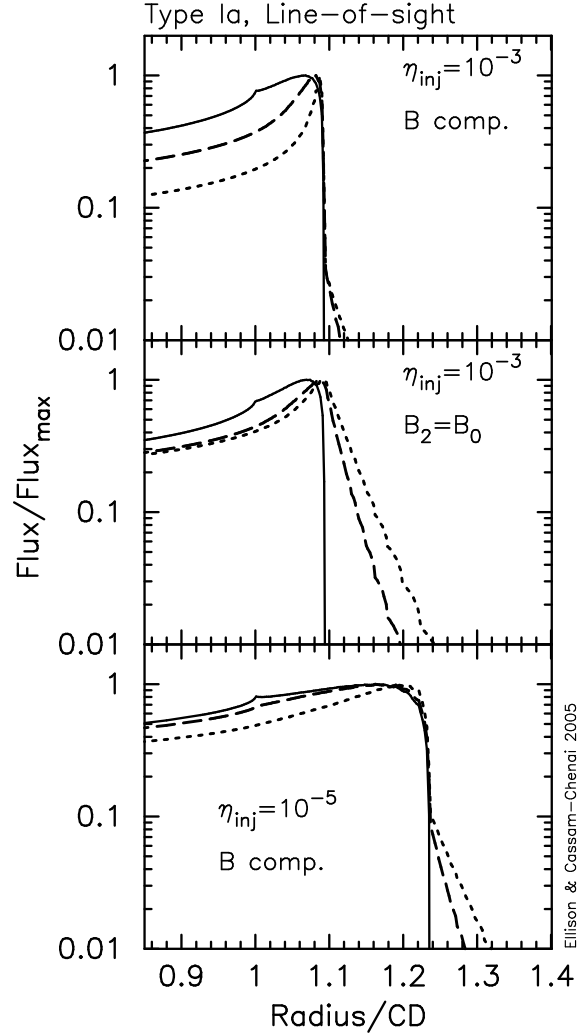}
\caption{Line-of-sight projections for the radial distributions with compressed $B$
  shown in Fig.~\ref{Ia_stack}. In all panels, the solid curve is radio (1--1.4 GHz),
  the dashed curve is low energy X-rays (0.1--1 keV), and the dotted curve is high
  energy X-rays (1--10 keV).
\label{Ia_LOS}}
\end{figure}

\begin{figure}         
\epsscale{.99}
\plotone{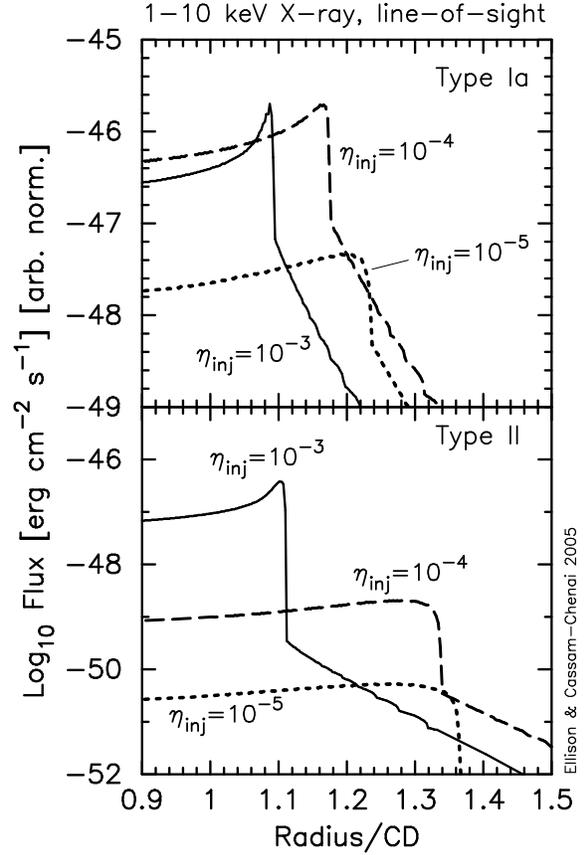}
\caption{Light-of-sight projections for the 1--10 keV X-ray band for various
  injection efficiencies as marked. All results include magnetic field compression
  and are for $\tSNR=400$\,yr.
  While the absolute normalization is arbitrary, the relative normalization between
  the various plots is correct (note the different vertical scales in the two panels).
\label{Ia_II_inj}}
\end{figure}

\begin{figure}         
\epsscale{.99} \plotone{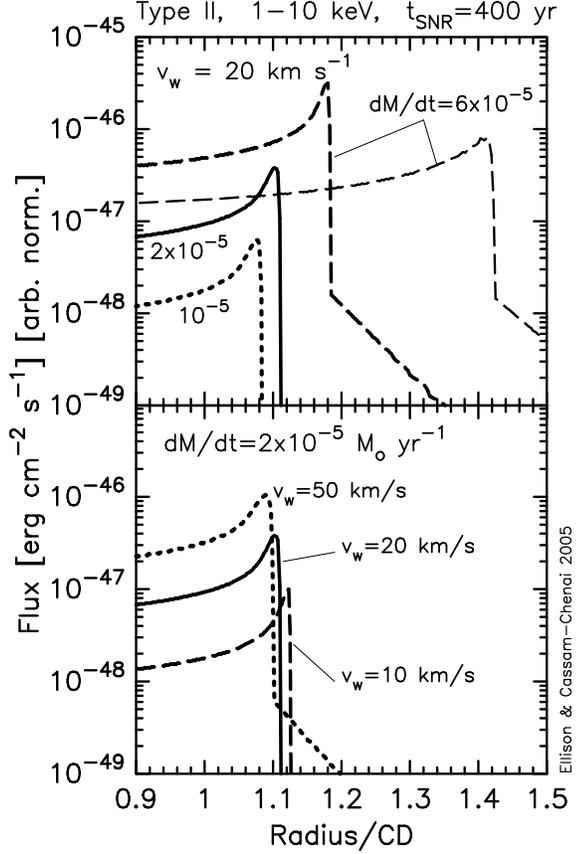} 
\caption{Line-of-sight projections for 1--10 keV
X-rays for our type II prototype with compressed magnetic field. The light-weight
dashed curve in the top panel has $\etainj=10^{-4}$, while all other models shown
have $\etainj=10^{-3}$. In the top panel, $\vWind=20$\,\kmps\ and $\Mdot$ (in units
of $\Msun$ yr$^{-1}$) is varied as indicated. In the bottom panel, $\Mdot= 10^{-5}
\,\Msun$ yr$^{-1}$ and $\vWind$ is varied.  The light-weight dashed curve in the top
panel with $\etainj=10^{-4}$ has an acceleration efficiency at $\tSNR=400$\,yr of
more than 70\% (similar to that shown in Fig.~\ref{CR_eff}) and demonstrates that
values of $\RfsRcd \sim 1.4$, as observed for Cas A, are consistent with efficient
DSA in type II \SNe.  As in Fig.~\ref{Ia_II_inj}, the absolute normalization is
arbitrary but the relative normalization between the various plots is correct.
\label{Mdot}}
\end{figure}

{We have emphasized here that another }
signature of efficient cosmic-ray ion production 
is the large reduction in the ratio of the radius of the forward shock to the radius
of the \CD, $\RfsRcd$.  If a large fraction of the shock energy goes into \rel\
particles and high-energy particles that escape from the shock system, $\Rtot \gg 4$
and the interaction region between the CD and FS will be denser and $\RfsRcd$ will be
smaller than with inefficient acceleration (Figs.~\ref{Ia_stack},
\ref{II_stack_10yr}, and \ref{Ia_II_inj}). This effect may explain observations of
$\RfsRcd \sim 1$ in Tycho's and Kepler's SNRs. Type II \SNe\ with pre-SN winds may
experience efficient DSA yet still show large $\RfsRcd \sim 1.3$--$1.4$, consistent
with observations of Cas A and 1E0102.2-7219 (Figs.~\ref{Ia_II_inj} and \ref{Mdot}).
While complicating factors such as an irregular ambient medium, dense knots, thin
sheets of emission, etc., exist in all SNRs, efficient DSA offers a natural
explanation for this important aspect of SNR morphology. Just as important, a
large value of $\RfsRcd$ observed in a type Ia SNR is strong evidence against
efficient DSA.

\begin{figure}         
\epsscale{.99}
\plotone{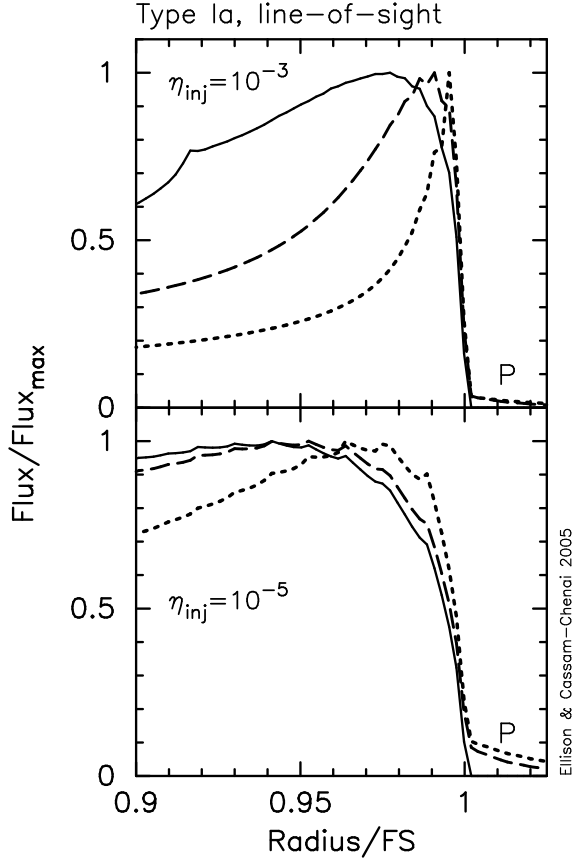}
\caption{Line-of-sight projections for the radial distributions shown in
  Fig.~\ref{Ia_stack} normalized to the forward shock radius. The magnetic field is
  compressed in both panels. As in the previous figures, the solid curve is radio
  (1--1.4 GHz), the dashed curve is low energy X-rays (0.1--1 keV), and the dotted
  curve is high energy X-rays (1--10 keV).  Note that the radio emission (solid
  curves) peaks well within the X-ray emission in all cases. The fluctuations, most
  noticeable in the radio emission for $\etainj=10^{-5}$, are numerical noise.
\label{Ia_LOS_FS}}
\end{figure}

\begin{figure}         
\epsscale{.99} \plotone{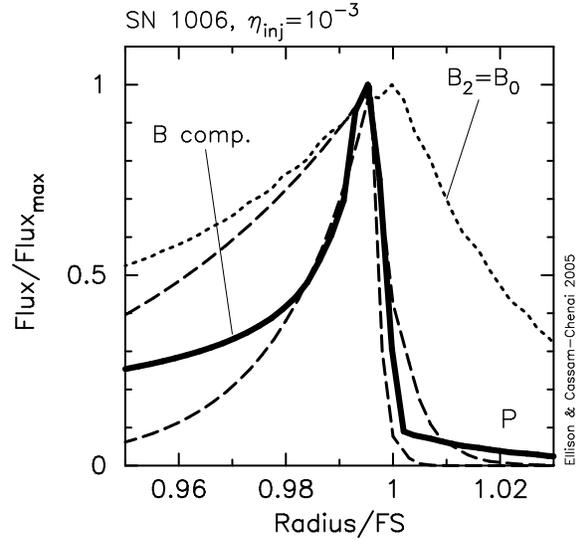} 
\caption{Comparison of X-ray line-of-sight
profiles from the CR-Hydro model with {\it Chandra} observations of SN 1006. The
dashed curves roughly span the maximum and minimum scale heights determined by
\citet{BambaEtal2003} where they assumed exponential profiles.  Using their Table~4,
we set the maximum (minimum) upstream scale height to be 3 (1) arcsec, and the
downstream maximum (minimum) scale height to be 30 (10) arcsec (the radius of SN 1006
is about $0.25^\circ$).  The solid curve is the X-ray
emission in the 1.2--2 keV band using our compressed $B$ model and for comparison, we
show (dotted curve) the 1.2--2 keV band without compressing the field. We have
positioned the peaks of the dashed curves to match the solid curve.
\label{Bamba}}
\end{figure}

{Yet another }
sign of efficient DSA is the presence of short scale
heights seen in nonthermal X-ray emission. Short scale heights are predicted with
efficient DSA because the shock will strongly compress the downstream magnetic field
and \syn\ losses will lower the emissivity immediately behind the FS. This results in
several related morphological effects.
First, thin sheets of X-ray emission (e.g., Fig.~\ref{Ia_II_inj}) should be common at
the FS, as is consistent with observations.
Second, projection effects should result in the distinct separation of the radio and
X-ray peaks (e.g., Fig.~\ref{Ia_LOS_FS}), also commonly observed.
Finally, as we show in
Fig.~\ref{Bamba}, the short scale heights seen in SN 1006
\citep[e.g.,][]{BambaEtal2003}, are most naturally explained as sharply peaked
emission behind the FS seen in projection \citep[][have already concluded this for SN
1006]{BKV2003}.
The actual upstream precursor has a long scale length, as expected for TeV electrons,
but is weak enough to avoid detection.

Supernova remnant SN 1006 seems a clear case where the efficient production of CR
ions is taking place, but remnants such as Tycho's and Kepler's, with $\RfsRcd \sim
1$, are also likely candidates.
The presence of a significant population of CR ions in young SNRs produces effects
that are readily observable in radiation produced by electrons and we have made
predictions, capable of being tested with {\it Chandra} and {\it
XMM-Newton}, to test this assertion.

\acknowledgments
The authors are grateful to A.~\textsc{Decourchelle} and J.~\textsc{Ballet} for the
discussions preceding this paper.  D.C.E. wishes to acknowledge the International
Space Science Institute (ISSI) in Bern, Switzerland for hosting a series of workshops
where some of the work presented here was done, as well as support from a NSF grant
(INT-0128883) and a NASA grant (ATP02-0042-0006).

\bibliographystyle{aa} 
\bibliography{c:/a_active/bibTeX/bib_DCE}

\end{document}